\documentclass[preprint,aps]{revtex4}

\usepackage{graphicx}
\usepackage{bm}
\usepackage{amsmath}
\usepackage{natbib}
\usepackage{dcolumn}

\def\bea {\begin{eqnarray}}
\def\eea {\end{eqnarray}}
\def\be {\begin{equation}}
\def\ee {\end{equation}}
\def\ben{\begin{enumerate}}
\def\een{\end{enumerate}}
\def\bi{\begin{itemize}}
\def\ei{\end{itemize}}

\def\hyphen{{\mbox{-}}}

\def\2p1h{2p\hyphen 1h }
\def\3p2h{3p\hyphen 2h }

\begin{document}




\title{Beta asymmetry parameter in the decay of $^{114}$In}


\author{F. Wauters}
\affiliation{Instituut voor Kern- en Stralingsfysica, Katholieke Universiteit Leuven,
B-3001 Leuven, Belgium}

\author{V. De Leebeeck}
\affiliation{Instituut voor Kern- en Stralingsfysica, Katholieke Universiteit Leuven,
B-3001 Leuven, Belgium}

\author{I. Kraev}
\affiliation{Instituut voor Kern- en Stralingsfysica, Katholieke Universiteit Leuven,
B-3001 Leuven, Belgium}

\author{M. Tandecki}
\affiliation{Instituut voor Kern- en Stralingsfysica, Katholieke Universiteit Leuven,
B-3001 Leuven, Belgium}

\author{E. Traykov}
\affiliation{Instituut voor Kern- en Stralingsfysica, Katholieke Universiteit Leuven,
B-3001 Leuven, Belgium}

\author{S. Van Gorp}
\affiliation{Instituut voor Kern- en Stralingsfysica, Katholieke Universiteit Leuven,
B-3001 Leuven, Belgium}

\author{D.~Z\'{a}kouck\'{y}}
\affiliation{Nuclear Physics Institute, Academy of Sciences of
Czech Republic, CZ-25068 \v{R}e\v{z}, Czech Republic}

\author{N. Severijns}
\affiliation{Instituut voor Kern- en Stralingsfysica, Katholieke Universiteit Leuven,
B-3001 Leuven, Belgium}

\date{\today}


\begin{abstract}
The $\beta$ asymmetry parameter $\widetilde{A}$ for the pure
Gamow-Teller decay of $^{114}$In is reported. The
low temperature nuclear orientation method was combined with a GEANT4
based simulation code allowing for the first time to address in detail
the effects of scattering and of the magnetic field.
The result, $\widetilde{A}$ = -0.994 $\pm$ 0.010$_{stat}$ $\pm$ 0.010$_{syst}$,
constitutes the most accurate value for the asymmetry
parameter of a nuclear $\beta$ transition to date. The value is in
agreement with the Standard Model prediction of $\widetilde{A}$ = -1 and
provides new limits on tensor type charged weak currents.

\end{abstract}

%
%

\pacs{23.40.Bw; 23.20.En; 24.80.+y; 27.60.+j}


\maketitle

The historic  measurement of the $\beta$ asymmetry parameter $A$
by Wu $\emph{et al.}$ ~\cite{wu57} strikingly showed the violation
of parity in weak interactions. Later this parameter was and still
is extensively studied in neutron decay, e.g. to test the unitarity of the quark
mixing matrix and to constrain the presence of right-handed
currents (for reviews see \cite{nico05,severijns06,abele08,amsler08}).
In nuclear decays only a few measurements of the $\beta$
asymmetry parameter were performed with the aim of testing symmetries of the weak interaction
and/or searching for physics beyond the Standard Model (see e.g.
\cite{boothroyd84,severijns06}). The most accurate result was
reported in Ref.~\cite{chirovsky80},
i.e. a (purely statistical) precision of 2~\% for the case of $^{60}$Co.
Still, such measurements can provide interesting information on
possible new physics \cite{jackson57}. The best choice are
fast (log~$ft$ $\leq$ 5) pure Gamow-Teller (GT)
transitions and the superallowed mirror transitions between
$T=1/2$ analog states \cite{naviliat91,severijns08}, for which
nuclear structure and other corrections contribute at most at the
permille level. $J^\pi \rightarrow J^\pi$ transitions
between non-analog states should be avoided as they usually contain
small isospin-forbidden Fermi contributions that originate from the
electromagnetic interaction and modify the value of $A$
(e.g. \cite{raman75,schuurmans00,severijns05}). For pure Fermi
transitions $A \equiv 0$.

Here we report a measurement of the asymmetry parameter for the
$1^+ \rightarrow 0^+$ pure GT
$\beta^-$~transition of $^{114}$In ($t_{1/2}$ = 72~s) with endpoint
energy of 1.989 MeV and log$ft$ = 4.473(5)~\cite{blachot02}.
Our method combines low temperature nuclear orientation
\cite{stone86} with GEANT4 simulations to address in detail, and
for the first time for this method, the
effects of scattering and of the magnetic field.
The result provides new information on tensor contributions to the
charged weak current. Other experiments are pursuing
similar goals \cite{flechard08,pitcairn09}.

The electron angular distribution for $\beta$ decay
of nuclei with vector polarization $\textbf{J}$ is
written as~\cite{jackson57}
\begin{eqnarray}
\label{eqn:jtw}  W \propto \left[ 1 + \frac{m }{E_e} b_{Fierz}+
\frac{\bf{p_e}}{E_e} \cdot \bf{J} \rm{A} \right]
\end{eqnarray}
\noindent with $E_e$ and $\bf{p_e}$ the total energy and momentum
of the $\beta$ particle and $m$ the electron rest mass.
Expressions for the Fierz interference term $b_{Fierz}$ and the
asymmetry parameter $A$ are given in Refs.~\cite{severijns06,jackson57}.
In the Standard Model
$b_{Fierz}$ = 0, while for a $J \rightarrow J-1$
pure GT transition, $A_{SM,GT}(\beta^\mp) = \mp 1$.
The observable actually determined by experiment is
$\widetilde{A} \equiv A / [ 1 + (m/\langle E_e \rangle)b_{Fierz} ]$.


A pure $^{114}$In sample  was obtained from the internal transition (IT) decay of
$^{114m}$In ($t_{1/2}$ = 49.5~d).
The latter was implanted at 70 kV with a dose of 1 $\times$ 10$^{12}$
at/cm$^2$ into a Fe foil (purity 99.99\%, thickness 50 $\mu$m).
The $\gamma$ spectrum showed no sign of contamination by another isotope.
%
%
%
%
The foil was soldered at 80 $^{\circ}$C (to prevent diffusion of
In in the Fe) on the sample holder of a
$^{3}$He-$^{4}$He dilution refrigerator. The latter served to polarize the
nuclei by cooling them to millikelvin temperatures in a strong magnetic hyperfine field
(in the plane of the foil) induced by a superconducting split-coil magnet.

%
%

The geometry was similar to the one shown in Fig.~7 of Ref.~\cite{wouters87}.
The $\beta$ particles were observed with two planar HPGe detectors \cite{wouters90}
with a sensitive diameter of 20 mm and thicknesses of 2 mm and
3 mm, respectively, placed at $0^\circ$ (axial detector) and
$90^\circ$ (equatorial detector) with respect to the vertical magnetic
field (i.e. orientation) axis. They were installed inside the 4 K radiation shield of
the refrigerator at (46~$\pm$~1) mm from the
sample, and operating at about 10 K with an
energy resolution of 4.5 keV. They were directly facing
the sample to minimize energy loss and
scattering effects. To further minimize the effect of scattering in the foil
(a major scattering component) the plane
of the foil was tilted $20^\circ$ towards the axial detector and
rotated towards the equatorial detector over $68^\circ$ (a compromise
to permit implantation of the beam).
The detectors were then at angles
of $20^\circ$ and $108.5^\circ$  with respect to the orientation axis.
To minimize also the effect of
the magnetic field on the $\beta$ particle trajectories the
measurements were performed in low external fields, i.e. $B_\mathrm{ext}
=$ 0.046~T, 0.093~T and 0.186 T.

The $\gamma$ rays of $^{114m}$In and of the $^{57}$Co{\it Fe}
nuclear orientation thermometer \cite{marshak86} were observed
with two large volume HPGe detectors. These were placed outside
the refrigerator at angles of $0^\circ$ and $90^\circ$ relative to
the magnetic field axis, at (70~$\pm$~2) mm from the sample.


The experimental angular distribution was determined as
\cite{krane86} $W(\theta) = N_{\textit{cold}}(\theta) /
N_{\textit{warm}}(\theta)$
%
%
\noindent with $\theta$ the angle with respect to the
magnetization (orientation) direction in the Fe foil and $N_{\it{cold,warm}} (\theta)$
the count rates in a given $\gamma$
peak, or energy bin in the $\beta$ spectrum, when the sample is
polarized (i.e. at millikelvin temperatures; \textit{cold}) or
unpolarized (i.e. at about 4.2 K; \textit{warm}). Note that
using a ratio of count rates reduces several systematic effects.




The experimental angular distribution of $\beta$ particles emitted
in allowed $\beta$ decay from polarized nuclei is \cite{krane86}
\begin{equation}
\label{eqn:W_beta} W(\theta) = 1 + f \frac{v}{c } \widetilde{A} P
Q \cos\theta \; ,
\end{equation}
\noindent with $v/c$ the initial velocity of the detected $\beta$
particles relative to the speed of light and $Q$ the solid angle
correction factor. Further, $P(J, \mu, T, B)$  is the degree of nuclear
polarization of the state with spin $J$ and magnetic moment $\mu$
at temperature $T$ in the polarizing magnetic field $B =
B_\mathrm{ext}+B_\mathrm{hf}$, with $B_{hf}$ = -28.7~T the internal hyperfine
field which the nuclei feel in the Fe foil. The
temperature of the sample which, via the Boltzmann distribution,
determines the degree of nuclear polarization is obtained from the
anisotropy of the 136 keV $\gamma$ ray of
the $^{57}$Co{\it Fe} thermometer.
Finally, the factor $f$ represents the fraction of
nuclei that feel the full orienting hyperfine interaction,
 $\mu B$. In the
two-site model used the fraction $(1 - f)$ is supposed to
feel no interaction at all.
%
%
The fraction $f$ was obtained from the anisotropy of the
$5^+ \rightarrow 1^+$ 190 keV IT of the implanted
$^{114m}$In. This was done for each field value as it was observed
\cite{schuurmans96} that part of the saturation magnetization of
the Fe foil obtained at 0.5 T was lost when the field was
reduced to a lower value for the measurements. Since emission of the 190 keV
$\gamma$ ray does not cause the $^{114}$In daughter
nuclei to recoil from their lattice position,
the values obtained for the fraction $f$ apply to $^{114}$In as well.

A GEANT4 based Monte-Carlo code \cite{kraev08} was used to
calculate the factor $Qcos\theta$ which includes the geometry of the setup,
the effect of the magnetic field on the $\beta$ particle
trajectories, and (back)scattering in the source, on the sample
holder and on the detector.
The geometrical $Qcos\theta$ values for zero magnetic field
and for the spectrum endpoint (where scattering effects are negligible)
were equal to $Qcos\theta$ = -0.930(6) and +0.314(2) for the axial
and equatorial detectors, respectively.

\begin{figure}[here]
\includegraphics[width=\columnwidth]{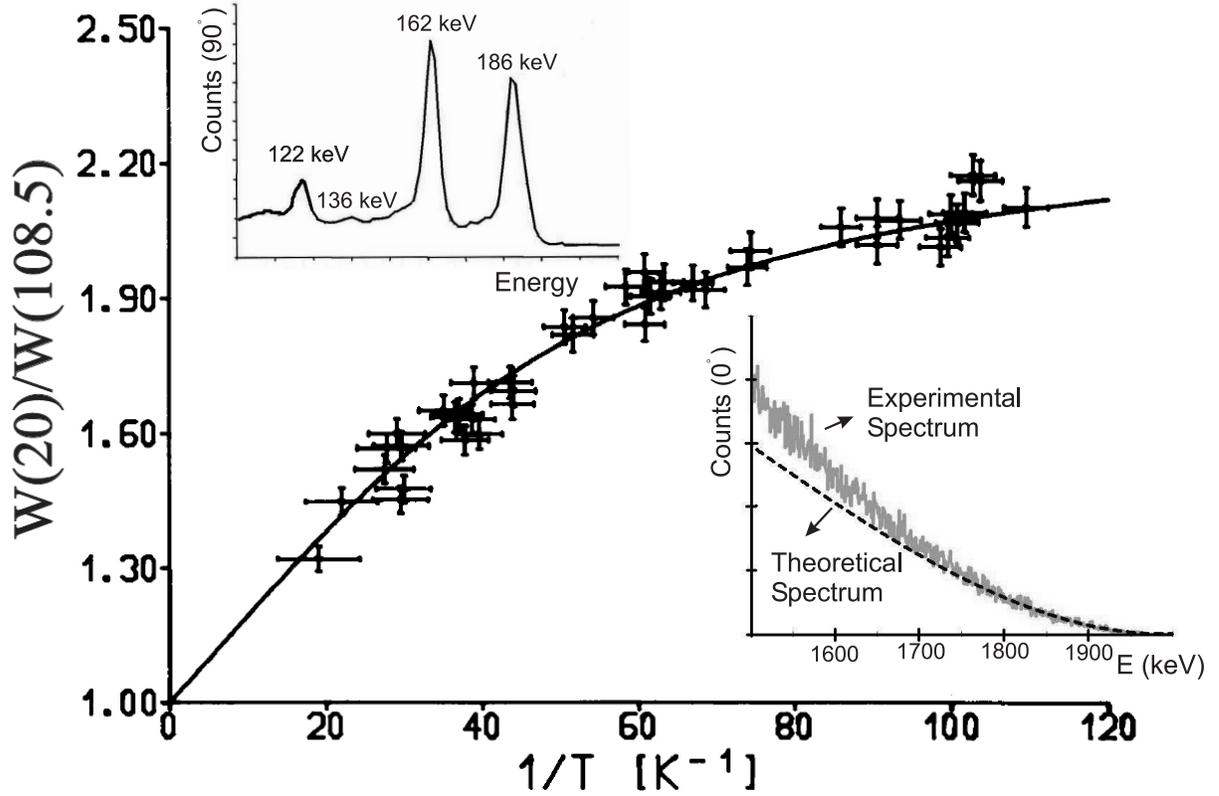}
\caption{\label{fig:beta-anisotropy} $\beta$ anisotropy of $^{114}$In
for the region between 1.700 and 1.830 MeV, in a field of 0.046 T ($\chi^2$/$\nu$ = 0.71).
Upper inset: Low energy part of the $\beta$ spectrum with $\gamma$ rays from $^{57}$Co and
conversion electrons from $^{114}$In. Lower inset: High energy part of the $\beta$ spectrum.}
\end{figure}

For the energy calibration the conversion electrons from $^{114}$In
(Fig.~\ref{fig:beta-anisotropy}) and the $\gamma$ rays from a $^{60}$Co
source were used.
The Ge detectors that were used in view of the high endpoint energy give rise to
rather significant scattering effects; the percentage of scattered
events was found to increase from 5~\% at about 1.75 MeV to
12~\% at about 1.60 MeV. Therefore, the analysis was limited
to the highest energy part of the $\beta$ spectrum where all disturbing
effects are minimal. The lower energy bound
for the region for analysis was then set at 1.700(10) MeV as simulations showed
that from this energy on the values of $Qcos\theta$ were affected by
less than 5~\% (relative) by the magnetic field, scattering, etc.
(Fig.~\ref{fig:beta-anisotropy}).
As the count rate for energies above 1.830(10) MeV was marginal the upper bound
for analysis was set at this value, such that $v/c$ = 0.9744(15).
Simulations showed the values of $Qcos\theta$ for the region from 1.700 MeV
to 1.830 MeV, and for the part of the spectrum above to be identical
(see Table~\ref{Table I}), indicating that the conditions for scattering
and magnetic field effects in the region used for analysis are very similar to
the ones at the endpoint.
The precision to which this holds for both detectors and the three magnetic
field values, was found to correspond to a 0.6~\% variation of
$\widetilde{A}$ that was subsequently assigned as a systematic error related
to the Monte-Carlo simulations. This turned out to be the largest
systematic error.

\begin{table}
\caption{\label{Table I} Differences of the simulated values of $Qcos\theta$
for the energy region from 1.700 MeV to 1.830 MeV used for analysis ('$\it{anal}$'),
and the region from 1.830 MeV up to the spectrum endpoint at 1.989 MeV ('$\it{end}$'),
for both detectors.}
\begin{ruledtabular}
\begin{tabular}{ccc}
    Magnetic field  & \multicolumn{2}{c}{$(Qcos\theta)^{anal} - (Qcos\theta)^{end}$}  \\
    \hline
                    &      axial detector      &   equatorial detector     \\
    \hline
         0.046 T    &         -0.0031(42)      &         -0.0032(25)       \\
         0.093 T    &         -0.0016(66)      &         +0.0039(37)       \\
\end{tabular}
\end{ruledtabular}
\end{table}

%
%
%
%
Fig.~\ref{fig:beta-anisotropy} shows the $\beta$ anisotropy
observed in a field of 0.046 T. Results for
$\widetilde{A}$ obtained from all three measurements are listed in
Table~\ref{tab:114In-results}. An overview of systematic
errors is given in Table~\ref{tab:114In-errors}:
The energy calibration led to a negligible systematic error.
The error related to the precision to which the geometry of the
setup and the magnetic field map that were used in the
simulations are known, was determined by repeating the analysis
with one standard deviation modified input for the simulations,
leading to systematic errors on $\widetilde{A}$ of 0.3~\% and
0.4~\%, respectively.
The magnetic field map was calculated from the magnet's dimensions
and properties provided by the manufacturer. The accuracy of it
was checked by a comparison with measured field values.
The difference between the calculated and measured values was then used as
error bar.
The precision to which the geometry of the $\gamma$ detection setup
(used to determine the fraction $f$) was known induced a 0.13~\%
systematic uncertainty on $\widetilde{A}$. The
experimental uncertainty on the magnetic hyperfine interaction
strength $\mu B$ used in the analysis translates into an uncertainty
in the degree of nuclear orientation. This induces a systematic error
of 0.16 \% from fitting the $\beta$ asymmetry, to obtain $f\widetilde{A}$,
and an additional 0.15 \% via the determination of $f$ from the 190 keV
$\gamma$ anisotropy. Since both errors are fully correlated they are added linearly.

Finally, analysis also accounted for the facts that
$^{114}$In, due to its short half-life, inherits a small part of the polarization
from its precursor, $^{114m}$In, and that some nuclei may not yet have reached
thermal equilibrium (i.e. the polarization corresponding to the sample temperature)
in the Fe lattice when they decay \cite{hagn81,klein86,venos03}.
As the $\mu B$ values for both isotopes
that govern these effects are precisely known
and, in addition, $\mu B$ for $^{114}$In is very large both effects are
small and can be fully accounted for \cite{severijns89,venos03}.
Varying then all relevant parameters within their error bars translated
into a variation of $\widetilde{A}$ of at most 0.3\%.

To account for the apparent field dependence of
$\widetilde{A}$ a systematic error of 0.4~\% was added,
determined by the average shift of the $\widetilde{A}$ values
required to get $\chi^2/\nu$ = 1.0.

Taking into account the statistical and systematic errors discussed above
(Tables~\ref{tab:114In-results} and \ref{tab:114In-errors}), our experimental
result is $\widetilde{A}$ = -0.990 $\pm$ 0.010(stat) $\pm$ 0.010(syst).


%
\begin{table}
\caption{\label{tab:114In-results}Results for $\widetilde{A}$.
The fraction $f$ was obtained from the anisotropy of the 190 keV
IT $\gamma$ ray of $^{114m}$In. Only statistical errors are listed
here. The error on $\widetilde{A}$ is a combination of the
statistics on the $\beta$ anisotropy (from which the product
$f \widetilde{A}$ was obtained) and the
statistical error on the fraction $f$ that
was used to extract $\widetilde{A}$. The error on the weighted
average of $\widetilde{A}$ was increased by a factor $\sqrt{\chi
^2/\nu}$ = 1.55. Systematic errors are listed in
Table~\ref{tab:114In-errors}.}
\begin{ruledtabular}
\begin{tabular}{cccc}
    $B_\mathrm{ext}$ [T] & fraction $f$ &  $\widetilde{A}$  \\
    \hline
    0.046  & 0.734(5)  &  -1.003(9)    \\
    0.093  & 0.803(8)  &  -0.987(13)    \\
    0.186  & 0.874(7)  &  -0.972(11)   \\
    \hline
    weighted average &  &  -0.990(10)
\end{tabular}
\end{ruledtabular}
\end{table}
\begin{table}[here]
\caption{\label{tab:114In-errors}Total error account}
\begin{ruledtabular}
\begin{tabular}{lcc}
    Systematic effect  & Correction  &  Error  \\
            &     [\%]    &   [\%]  \\
    \hline
    Energy calibration                                      &        &  0.005    \\
    Simulations of the $\beta$ asymmetry ($Q cos\theta$)    &        &  0.60    \\
    Geometry of the $\beta$ detection setup                 &        &  0.3     \\
    Magnetic field map                                      &        &  0.4     \\
    Apparent magnetic field dependence                      &        &  0.4    \\
    Accuracy in setting the magnetic field                  &        &  0.001   \\
    Geometry of the $\gamma$ detection setup ($f$)          &        &  0.13    \\
    Effect of error on $\mu B$ on the values of            &        &          \\
    ~ ~ $f$ (0.15 \%) and $f\widetilde{A}$ (0.16 \%)         &        &  0.31    \\
    Possible incomplete relaxation of $^{114}$In and        &        &          \\
    ~ ~ polarization inherited from $^{114m}$In             &        &  0.3     \\
    Induced form factors $b/Ac$ and $d/Ac$                  & -0.41  &  0.30    \\
\hline
    Systematics total                                       & -0.41  &  1.0     \\
    Statistics (see Table~\ref{tab:114In-results})          &        &  1.0     \\
\hline
    Sum                                                     & -0.41  &  1.4     \\

\end{tabular}
\end{ruledtabular}
\end{table}
%


At this level of precision a small correction for the effect of
recoil order terms \cite{holstein74} is necessary. Since we deal
with a $1^+ \rightarrow 0^+$ pure GT transition only the weak magnetism,
$b$ in the notation of Holstein \cite{holstein74}, and the first class
induced tensor term, $d$, are important. Further, from systematics
(see e.g. \cite{calaprice76,deleebeeck09}) we estimate
$b/Ac=4.6\pm2.7$ and $d/Ac=\pm(1.2\pm1.2)$ with $A$ here being the mass
of the nucleus and $c$ the GT form factor \cite{holstein74}.
One then calculates a recoil correction of $-0.0041(30)$ which leads
to $\widetilde{A}$ = -0.994 $\pm$ 0.010(stat) $\pm$ 0.010(syst).

\begin{figure}[here]
\includegraphics[width=\columnwidth]{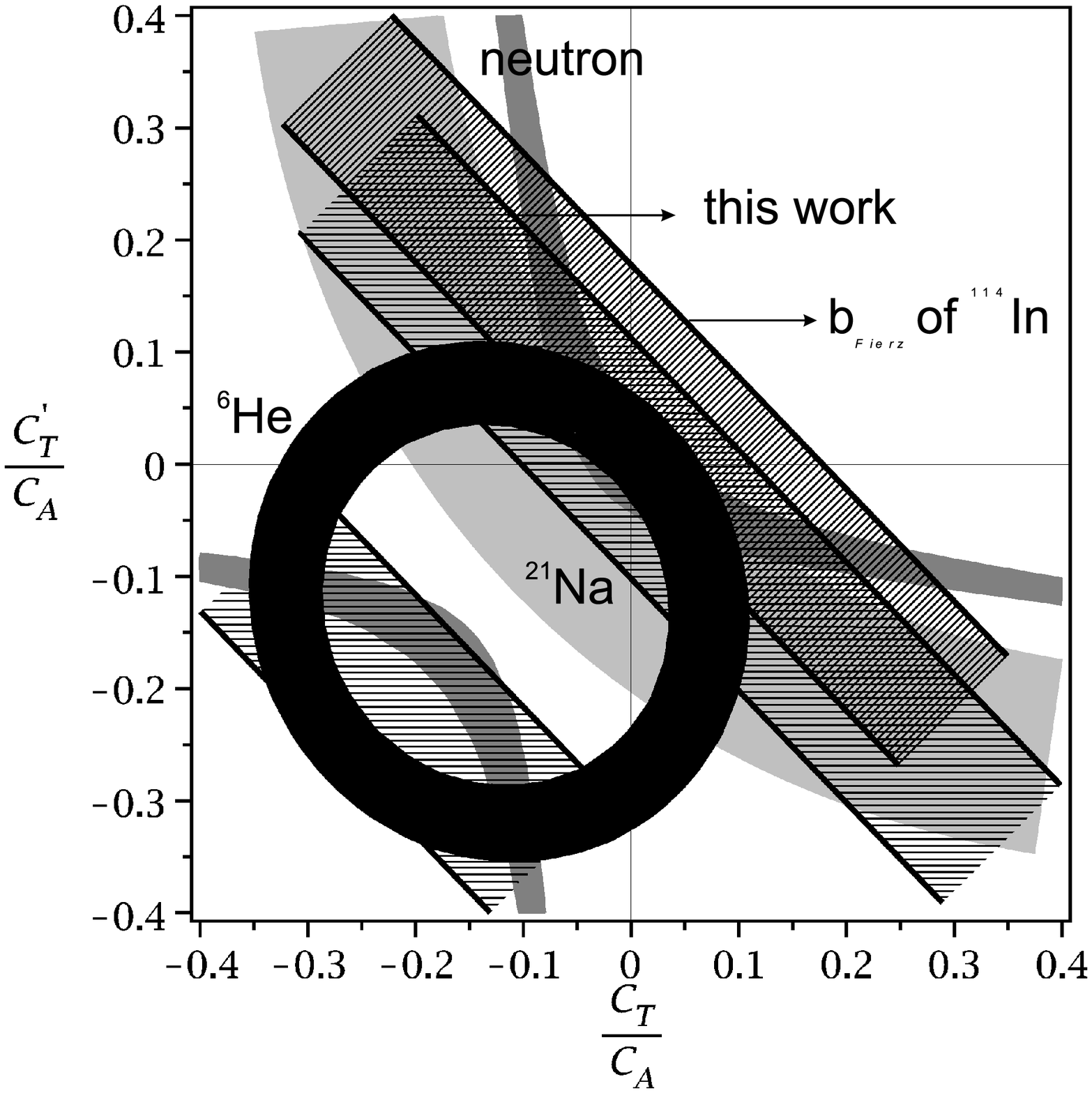}
\caption{\label{fig:tensor-constraints} Limits (90\% C.L.) on
time reversal invariant tensor couplings from correlation measurements
in nuclear $\beta$ decay: the Fierz interference term $b_{Fierz}$
from a spectrum shape measurement for $^{114}$In (with only
statistical errors quoted) \cite{daniel61,daniel64},
the $\beta$ asymmetry parameter $\widetilde{A}$ in the decay of
$^{114}$In (this work) and in free neutron decay \cite{amsler08},
and the $\beta$-$\nu$ correlation coefficient $a$ in the decays
of $^6He$ \cite{johnson63,gluck98} and $^{21}$Na \cite{vetter08}
(see also ref.~\cite{pitcairn09}). Limits for the
mixed Fermi/GT transitions of the neutron and $^{21}$Na
are for scalar couplings $C_S = C^\prime_S = 0$.}
\end{figure}

Being in agreement with the Standard Model value
$\widetilde{A}_{SM}^{\beta^-} = -1$ this result can now be used to
constrain physics beyond.
%
%
Our value corresponds to a lower limit of $M_2$ = 230 GeV/c$^2$ (90\% C.L.)
for the mass of the weak boson eigenstate $W_2$ that is mainly related to a $W_R$ boson that couples
to right-handed neutrinos \cite{beg77}. This is less stringent than limits
from other experiments in $\beta$ decay
\cite{severijns93,severijns98,allet96,thomas01,serebrov98,schumann07}.
However, our result is of interest for tensor type charged weak currents.
Assuming maximal parity violation and time reversal invariance for
vector and axial-vector currents, one has
for a $J \rightarrow J-1$ pure GT $\beta^-$ transition \cite{jackson57}:
\begin{eqnarray}
%
\label{eqn:Aparapprox_exot}
%
\widetilde{A}_{GT}^{\beta^-} & \simeq & A_{SM} + \frac{ \alpha Z m } {p_e} Im \left( \frac{ C_T + C^{\prime}_T }{ C_A } \right) \nonumber \\
& & + \frac{\gamma m} {\langle E_e \rangle} Re \left(\frac{ C_T + C^{\prime}_T }{ C_A } \right) + Re \left( \frac{ C_T C^{\prime *}_T } { C_A^2 } \right) \nonumber \\
& & + \frac{ |C_T|^2 + |C^{\prime}_T|^2 }{ 2 C_A^2 }
\end{eqnarray}
%
%

\noindent with $\alpha$ the fine-structure constant, $Z$ the
charge of the daughter nucleus, $\gamma = \sqrt{ 1 - (\alpha
Z)^2}$ and $C^{(\prime)}_{V,A,T}$ coupling constants for
vector, axial vector and tensor interactions. Primed (unprimed) coupling constants are for
parity violating (invariant) interactions. The term with
$\gamma m / \langle E_e \rangle$ comes from the Fierz interference term $b_{Fierz}$
in Eq.~(\ref{eqn:jtw}).

With an average kinetic energy of about 1.76 MeV for the observed
$\beta$ particles, the sensitivity factors in
Eq.(\ref{eqn:Aparapprox_exot}) are $\alpha Z m/ \langle p_e \rangle = 0.084$
and $\gamma m / \langle E_e \rangle = 0.209$. Assuming time reversal invariance,
i.e. $Im[(C_T + C^{\prime}_T)/C_A]$ = 0,
as was verified at the level of about 1\% in a measurement with $^8$Li
\cite{huber03}, and to first order in $C_T$, Eq.~(\ref{eqn:Aparapprox_exot})
yields for time reversal invariant, real tensor couplings
(i.e. the Fierz interference term)
$-0.082 < (C_T+C_T^\prime)/C_A < 0.139$ (90 \%C.L.).
In Fig.~\ref{fig:tensor-constraints} this result is compared to
limits from other experiments in nuclear $\beta$ decay.

The most accurate measurement of the $\beta$ asymmetry parameter for a 
nuclear $\beta$ transition to date was reported. Crucial to this 
was the use of a GEANT based simulation code for this type of measurements.  
Our result provides limits for time reversal invariant tensor couplings
in the weak interaction that are competitive with those from other
experiments. A still higher sensitivity can be obtained if $\beta^-$ transitions
with a lower endpoint energy are observed. E.g. for $E_e \simeq$ 600 keV
one has $\gamma m / E_e \simeq 0.4$ which yields, for the same accuracy,
two times more stringent limits.
Such measurements are in progress.

This work was supported by the Fund for Scientific Research
Flanders (FWO), project GOA/2004/03 of the K.U.Leuven and the grant LA08015
of the Ministry of Education of the Czech Republic.



%






\end{document}